\documentclass[nofootinbib,aps,plb,twocolumn,showpacs,superscriptaddress,groupedaddress,longbibliography]{revtex4-1}
\usepackage[fleqn]{amsmath}
\usepackage[T1]{fontenc}
\usepackage[latin9]{inputenc}
\usepackage{amstext}
\usepackage{amssymb}
\usepackage{braket}
\usepackage{graphicx}
\usepackage{hyperref}
\usepackage{natbib}

\begin{document}

\author{Karim Ghorbani}
 \email{karim1.ghorbani@gmail.com}
\affiliation{Physics Department, Faculty of Sciences, Arak University, Arak 38156-8-8349, Iran}
\author{Parsa Hossein Ghorbani}
 \email{parsaghorbani@gmail.com}

\title{Strongly First-Order Phase Transition in Real Singlet Scalar Dark Matter Model}
\affiliation{Applied Physics Inc., Center for Cosmological Research, \\
 300 Park Avenue, New York, NY 10022, USA}

\begin{abstract}

The extension of the standard model by a real gauge singlet scalar is the simplest but the most studied model
with sometimes controversial ideas on the ability of the model to address the dark matter and the electroweak phase transition issues simultaneously. 
For this model, we obtain analytically slightly different conditions for strongly first-order electroweak phase transition and 
apply that in computation of the dark matter relic density where the real scalar plays the role of the dark matter particle. 
We show that the scalar in this model before imposing the invisible Higgs decay constraint, can be responsible 
for all or part of the dark matter abundance, while at the same time gives rise 
to a strongly first-order electroweak phase transition required for the baryogenesis. 
When the constraints from the direct detection experiments such as XENON100 or LUX/XENON1t are considered, the model is excluded completely.

\end{abstract}


\maketitle

\section{Introduction}\label{in}

Apart from the discovery of the Higgs particle -the last elementary particle and the first scalar field discovered in nature- 
at the Large Hadron Collider (LHC) 
\cite{Aad:2012tfa,Chatrchyan:2012xdj}, 
the search for a footprint of beyond the Standard Model (BSM) of elementary particles 
by the experiments at the LHC has 
ended up to null results so far \cite{Baak:2012kk}. However, there are strong motivations 
to look for beyond the standard model. 
Some examples are
the absence of a well-established mechanism on how the Higgs gets a non-zero vacuum expectation value ({\it vev}) in the early universe, 
the matter-antimatter asymmetry observed today in the universe, and the mystery 
of dark matter (DM), which in all cases the existence of at least one more degree of freedom in the 
SM seems inevitable. 
On the other hand, the Higgs might not be the only scalar field in nature 
and the existence of further scalar degrees 
of freedom is not unlikely. The first possible addition of such a scalar would be 
the standard model (SM) plus a gauge-singlet scalar field. 
This is the simplest model beyond 
the standard model and has been studied vastly from various aspects in the past.  
These investigations may be divided in two categories. Once when the scalar field is stable under a $\mathbb{Z}_2$ discrete symmetry
and gets employed solely to explain the observed DM abundance and is constrained by direct and indirect searches, 
\cite{Silveira:1985rk,McDonald:1993ex,McDonald:2001vt,Bento:2000ah,Burgess:2000yq,He:2008qm,Lerner:2009xg,Gonderinger:2009jp,Farina:2009ez,Guo:2010hq,Profumo:2010kp, Biswas:2011td,Djouadi:2012zc,Feng:2014vea, Profumo:2014opa,Duerr:2015aka,Han:2016gyy,Sage:2016xkb,Wu:2016mbe}, 
or alternatively when the extra singlet scalar is used to give a first-order electroweak phase transition (EWPT) required by the Baryogenesis
\cite{Espinosa:1993bs,Choi:1993cv,Brustein:1998du,Ham:2004cf,Ahriche:2007jp,Profumo:2007wc,Das:2009ue,Espinosa:2011ax,Fuyuto:2014yia,Kurup:2017dzf}.
In the latter case, the singlet scalar is not necessarily required to get zero {\it vev} after the EWPT. 

In the most recent work on the status 
of the singlet scalar dark matter by the GAMBIT collaboration \cite{Athron:2017kgt}, the Bayesian and frequentist global fits
on the nuisance parameters, after imposing the constraints from 
the Planck for the relic density, the LUX, PandaX, SuperCDMS, XENON100 in direct searches, the invisible Higgs decay at the 
LHC, the IceCube for the DM annihilation to high energy neutrinos in the Sun, and the {\it Fermi}-LAT for DM annihilation 
into gamma rays in dwarf galaxies, 
show that the viable DM mass in the singlet scalar model
sits either in the range $\sim 125-300$ GeV or is about $\sim1$ TeV if the scalar constitutes all the dark matter abundance. 
A global fit of the $\gamma$-ray galactic center excess in the real singlet scalar dark matter model is accomplished in Ref. 
\cite{Cuoco:2016jqt}. These status reports on singlet scalar model have been only on DM part. 

There are a few works that have addressed separately or simultaneously both the DM and the EWPT
in the singlet scalar model \cite{Barger:2007im,Cline:2012hg,Alanne:2014bra,Vaskonen:2016yiu,Beniwal:2017eik}.
For instance,  in
Ref. \cite{Cline:2012hg} it is argued that the scalar can be responsible for only $3\%$ of the dark matter 
relic abundance and to give a first-order EWPT while evading the XENON100 direct detection (DD) bound. Ref. \cite{Vaskonen:2016yiu} however reports a slightly different result from \cite{Cline:2012hg} arguing that the model is completely unable to account for both the DM relic density and a first-order EWPT without considering the constraint from the DD experiments. 

In this paper we scrutinize the question of the first-order EWPT in the real singlet scalar model and realize that the previous literature use {\em unnecessary} stronger first-order conditions which make the results slightly different. We elaborate this point in the paper and analytically compute the first-order EWPT conditions. Then we use the {\tt micrOMEGAs} package for embedding both 
problems of the DM relic density and the electroweak EWPT in one code
to explore the viable regions of the parameter space. We also impose the washout criterion as a requirement for the phase transition to be strongly enough for the baryogenesis. Finally we update the direct detection constraint to the recent results from 
LUX and XENON1t. 

The paper is organized as the following. In Sec. \ref{phase} we introduce the model and obtain analytically the first-order phase transition conditions and the
washout criterion, then in Sec. \ref{dark} we numerically compute the dark 
matter relic density and the DM-nucleon cross section while imposing the strongly first-order phase transition.
We summarize the results in Sec. \ref{conc}.

\section{First-Order Phase Transition}\label{phase}

The model is the simplest renormalizable extension of the SM with an additional real singlet scalar denoted by $s$ here. 
The real scalar $s$ 
interacts with the SM through the Higgs portal having a quadratic interaction with the Higgs particle. Therefore, beside the 
Higgs and the scalar potentials, 
\begin{equation}\label{vhvs}
\begin{split}
& V_H= -\mu_h^2 H^\dagger H  +  \lambda_h (H^\dagger H)^2\\
& V_s= -\frac{1}{2}\mu_s^2 s^2  + \frac{1}{4}\lambda_s s^4 \,,
 \end{split}
\end{equation}
the total potential includes also the interaction part, 
\begin{equation}\label{vint}
 V_{\text{int}}= \lambda_{hs} \, s^2 H^\dagger H \,,
\end{equation}
where $H$ denotes the Higgs $SU(2)$ doublet and $\lambda_{hs}$ is the scalar-Higgs interaction coupling
playing an important role in our analysis. There are only two free parameters 
in the model, i.e. the scalar mass and the coupling, $\lambda_{hs}$. Note that we are not considering 
the terms in the potential that do not respect the $\mathbb Z_2$ discrete symmetry, such that  
the scalar can be a stable dark matter particle in the so-called {\it freeze-out} mechanism. 
Gauging away three components of the Higgs doublet, only one real component, $h$, remains and the Higgs field in 
eqs. (\ref{vhvs}) and (\ref{vint}) can be replaced by $H^\dagger=\frac{1}{\sqrt{2}}(0~h+v_h)$ after the electroweak symmetry breaking. 
When the temperature is very high the theory consisting of the SM and the new real scalar, lives in its symmetric phase. 
In this energy scale, the Higgs takes zero vacuum expectation value while the scalar could have zero or non-zero {\it vev}. 
\footnote{
Note that in the early universe, the dark matter freezes out from plasma of particles in a temperature 
much lower than the electroweak phase transition 
temperature, $T_c$. Therefore, to be more accurate, it only suffices that the real scalar takes zero {\it vev} just above the freeze-out temperature, $T_f$.}
The tree-level the total potential at high temperature can then be written as, 
\begin{equation}\label{vtr}
 V_{\text{tr}}(h,s)=-\frac{1}{2}\mu_{h}^{2}h^{2}-\frac{1}{2}\mu_{s}^{2}s^{2}+\frac{1}{4}\lambda_{h}h^{4}
+\frac{1}{4}\lambda_{s}s^{4}+\frac{1}{2}\lambda_{hs} s^{2}h^{2}\,. 
\end{equation}
As the universe cools down, at the time of electroweak symmetry breaking at lower temperatures, 
we assume a one-step phase transition that the {\it vev}s of the scalars, $h$ and $s$, change from 
$(v_0 =0, w_0 \neq 0 )$ to $(v\neq 0,w = 0 )$ at temperature $T_c$
where $v \equiv \braket{h}$ and $w \equiv \braket{s}$\footnote{At very high temperature the {\it vev} of the scalar must be vanishing, as can be seen easily from the thermal effective potential. However at intermediate temperatures the $w_0\neq 0$ can be a deeper minimum. We ignore the transition from $w_0=0$ to $w_0\neq 0$ which does not change the nature of EWPT.}. We require that after the phase transition $w=0$ because otherwise 
the $\mathbb Z_2$ symmetry is broken and the scalar $s$ can no longer be taken as the dark matter candidate.

Along the lines of \cite{Espinosa:2011ax}, in addition to the tree-level barrier we include also the 
one-loop thermal potential in order to get a strong EWPT, 
\begin{equation}\label{vt}
 V_T^{\text{1-loop}}(h,s;T)\simeq  \left( \frac{1}{2} c_h h^2 + \frac{1}{2} c_s s^2 \right) T^2 \,,
\end{equation}
where the parameters $c_h$ and $c_s$ are the following, 
\begin{equation}
\begin{split}
 &c_h=\frac{1}{48}\left( 9g^2 + 3g'^2 + 12y_t^2 + 12\lambda_h + 4\lambda_{hs} \right) \\
 &c_s= \frac{1}{12}\left( \lambda_{hs} + 3\lambda_s  \right)\,.
 \end{split}
\end{equation}
The one-loop correction at zero-temperature in the effective potential as pointed out in \cite{Ghorbani:2017jls,Ghorbani:2017lyk} is negligible,
therefore the thermal one-loop effective potential can approximately be written as, 
\begin{equation}\label{veff}
 V_{\text{eff}}(h,s;T)=V_{\text{tr}}(h,s)+V_T^{\text{1-loop}}(h,s;T)\,.
\end{equation}
The critical temperature $T_c$ of the electroweak phase transition, is the temperature at which
the free energy (thermal effective potential in eq. (\ref{veff})) in the symmetric phase equals 
the free energy in the broken phase. In other words, the thermal effective potential gets two degenerate
minina at $T=T_c$. Note the fact that we deal with two types of symmetries here. One is the $\mathbb{Z}_2$ discrete symmetry in $s$ which must 
exist below the freeze-out temperature, $T_f$, and the $SU(2)$ electroweak symmetry which is unbroken in high temperatures.
The phase transition process we consider here is a transition from 
$(\braket{h},\braket{s})=(v_\text{sym}=0,w_\text{brk})$ to $(\braket{h},\braket{s})=(v_\text{brk},w_\text{sym}=0)$. We recall that 
at temperatures above $T_c$, the scalar field $s$,  can always have non-zero vacuum expectation value. 
The minima of the thermal effective potential in eq. (\ref{veff}) read, 
\begin{equation}\label{vev1}
\begin{split}
v_{\text{sym}}=0 ~~\text{and} ~~v_{\text{brk}}^{2}\left(T\right)
=\frac{\mu_{h}^{2}-c_{h}T^{2}}{\lambda_{h}}\equiv \frac{\mu_h^2(T)}{\lambda_h}\,,
\end{split}
\end{equation}
where $v_{\text{brk}}$ is the $T$-dependent Higgs {\it vev}, and 
\begin{equation}\label{vev2}
\begin{split}
w_{\text{sym}}=0~~\text{and}~~ w_{\text{brk}}^{2}\left(T\right)=\frac{\mu_{s}^{2}-c_{s}T^{2}}{\lambda_{s}} \equiv \frac{\mu_s^2(T)}{\lambda_s}\,,
\end{split}
\end{equation}
with $w_{\text{brk}}$ being the $T$-dependent {\it vev} of the scalar $s$. 

\begin{figure}
 \includegraphics[angle=-90, scale=.32]{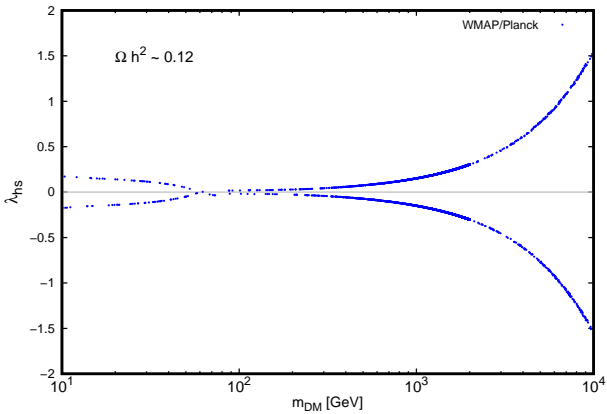}
  \caption{The plot shows the viable range of the coupling $\lambda_{hs}$ and the DM mass which gives rise to the correct
  relic density for the DM in the real singlet scalar model.}
  \label{relic}
\end{figure}

Now the critical temperature below which the transition from $(v_\text{sym}=0,w_\text{brk})$
to $(v_\text{brk},w_\text{sym}=0)$ takes place is obtained by solving 
$V_{\text{eff}}(0,w_\text{brk};T_c)=V_{\text{eff}}(v_\text{brk},0;T_c)$. 
The answer can be expressed as,  
\begin{equation}\label{Tc}
 T_c=\sqrt{\frac{ a - b}{c}}\,,
\end{equation}
with the parameters $a, b$ and $c$ defined as, 
\begin{equation}\label{abc}
 \begin{split}
  a=&   c_h \mu_h^2  \lambda_{s}  -c_s \mu_s^2 \lambda_h
  \,, \\
  b=&  \mid  c_h \mu_s^2  - c_s \mu_h^2  \mid
  \sqrt{\lambda_h \lambda_s }\,, \\
  c=& c_s^2 \lambda_h - c_h^2 \lambda_s\,.
 \end{split}
\end{equation}
At temperatures lower than the critical temperature, $T< T_c$ the minimum at $(v(T)= \sqrt{\mu_h^2(T)/\lambda_h},0)$ must be the global minimum down to $T=0$. This means that the $T^2$-derivative of  $\Delta V_{\text{brk-sym}}(T)$, 
 \begin{equation}
  \Delta V_{\text{brk-sym}}(T)=V(0,w(T))-V(v(T),0) >0 \,,
 \end{equation}
 must satisfy,
 \begin{equation}
  \frac{d\Delta V_{\text{brk-sym}}(T)}{dT^2}\Big\vert_{T=T_c} < 0\,,
 \end{equation}
 which leads to, 
 \begin{equation}\label{globmin}
  c_h \sqrt{\frac{\lambda_s}{\lambda_h}} > c_s\,.
 \end{equation}

It is assumed in the literature (see e.g. \cite{Vaskonen:2016yiu}) that the minimum $(0,w_0)$  must exist for all temperatures from $T=0$ to very large $T$, say $T\to \infty$. This puts strong constraints on $c_s$ and $\mu_s^2$: $c_s<0$ and $\mu_s^2>0$, which we explain in a moment that are not the correct conditions. 
On the other hand, the minimum $(v,0)$ should coexist together with the minimum $(0,w_0)$ for temperatures $T\leq T_c$, while the minimum $(v,0)$ must remain the deepest minima until $T=0$. The latter condition is given by Eq. (\ref{globmin}). The global minimum condition for $(v,0)$ given in Eq. (\ref{globmin}) and the (not very correct) local minimum conditions for $(0,w_0)$ mentioned above, have no overlap in the space of parameter, hence it has been inferred in \cite{Vaskonen:2016yiu} that the EWPT cannot be of first-order. 

Now we explain why the local minimum condition considered e.g. in \cite{Vaskonen:2016yiu} does not necessarily hold. Although, the local minimum condition at $(0,w_0)$ for $T \in [0,\infty)$ is sufficient, but it is not necessary. In fact, it is enough that the local minimum $(0,w_0)$, exist for even a short time before the EWPT, say from $T_c + \delta T $ to $T_c$ with $\delta T $ an arbitrary small value and $T_c$ being the critical temperature. Above the temperature $ T_c + \delta T $, the minimum may be either $(0,w_0)$ or $(0,0)$. Any possible change in $w_0$ above $T_c+\delta T$, has no effect on the electroweak phase transition. Considering this fact and in the limit $\delta T \to 0$, the local minimum condition for $(0,w_0)$ becomes,
\begin{equation}\label{localmincon}
 c_s< \frac{\mu_s^2}{T_c^2}\,.
\end{equation}

One of the Sakharov's condition for the baryon asymmetry is guaranteed by the suppressed sphaleron rate in the Higgs broken phase. 
This is equivalent to the condition $v_c/T_c>1$ that is called the {\it washout criterion} in which $v_c \equiv v_{\text{brk}}(T_c)$. 
For the solution we found in eq. (\ref{Tc}) we have therefore the following condition, 
\begin{equation}\label{vctc}
 \frac{v_c(T_c)}{T_c}>1\,
\end{equation}
with, 
\begin{equation}\label{vpm}
 v_c=\frac{c_s \mu_h^2-c_h \mu_s^2}{c_s \sqrt{\lambda_h} - c_h \sqrt{\lambda_s}}\,.
\end{equation}

\begin{figure}
 \includegraphics[angle=-90,scale=.32]{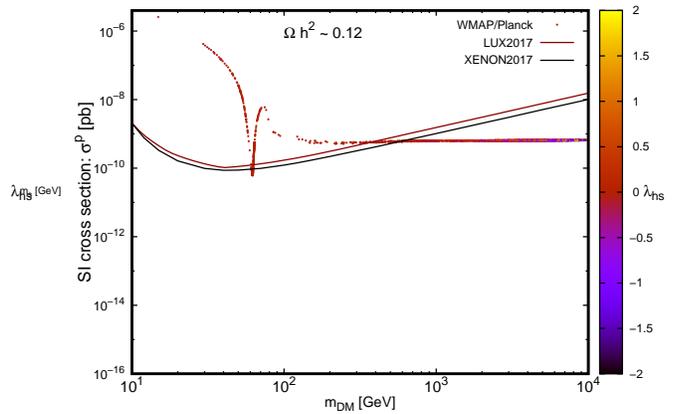}
 \caption{The relic density viable space saturates the spin-independent DM-nucleon cross section limit provided by 
 LUX/XENON1t at DM mass of $\sim 600$ GeV.}
  \label{dir}
\end{figure}

The stability conditions at $T=0$ still reduces the number of independent parameters. The Higgs physical mass fixes the 
parameter $\mu_h^2$ as $\mu_h^2=m_H^2/2$ with $m_H=125$ GeV. The Higgs quartic coupling is also fixed as 
$\lambda_h=m_H^2/2v_h^2\simeq 0.129$. Then from eq. (\ref{abc}), $\lambda_s>0$. The physical mass of the scalar $s$ (dark matter mass) 
is given by 
 $m_\text{DM}^2\equiv m_s^2=-\mu_s^2+\lambda_{hs}v_h^2$.
The positivity of the dark matter mass then requires $\mu_s^2<\lambda_{hs} v_h^2$.

\section{Dark Matter}\label{dark}
Another important issue we take into account in the simple model of the real scalar extension of the SM is the problem of the dark matter. 
Because of the $\mathbb{Z}_2$ discrete symmetry on the scalar $s$ we considered in eq. (\ref{vtr}), the scalar is stable
and is taken as the {\it weakly interacting massive particle} (WIMP). The DM particle is in thermal 
equilibrium with the SM particles in the early universe, but it detaches from other particles at the freeze-out temperature, $T_f$ after the universe 
is expanded enough with the Hubble rate. 
The {\it vev} of the Higgs particle at zero temperature is $v_h=246$ GeV and for 
the singlet scalar is $v_s=0$, hence there is no mixing between the Higgs and the DM particle. The only annihilation channel we deal with in this model is 
$ss\to \text {Higgs}\to \text{SM}$, therefore the only independent coupling in the Lagrangian that enters in the dark matter annihilation process and the dark matter elastic scattering off the nuclei is the $\lambda_{hs}$ in eq. (\ref{vtr}). 
Other parameters
in the theory does not affect the computation of the relic density and the DM-nucleon elastic scattering cross section. Nevertheless, 
they are present in the EWPT conditions as was discussed in the previous section.  
In our computations, we confirm that the EWPT occurs before
the dark matter freezes-out from the thermal equilibrium. This means that at the time of freeze-out the Higgs is already in a non-zero {\it vev} or equivalently the theory is already in its broken phase (the Higgs and the SM fermions are massive) 
and the singlet scalar has got the $\mathbb{Z}_2$ symmetry. 

\begin{figure}\label{dmdir}
    \includegraphics[angle=-90,scale=.32]{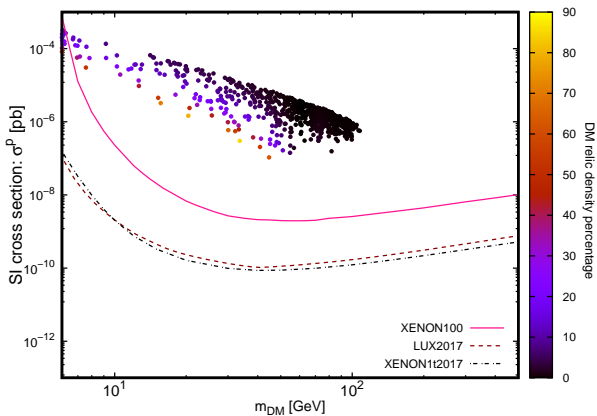}
      \caption{The DM mass against the DM-nucleon cross section when the electroweak phase transition is strongly first-order 
      and the singlet scalar constitutes $0-100\%$ of the total DM relic density 
      $\Omega_{\text{DM}} h^2 \sim 0.12$. The DM relic density and the strongly EWPT can be explained simultaneously by singlet scalar model with DM mass smaller that $200$ GeV. The XENON100 cross section bound excludes almost all the parameter space except for a small DM mass range $6-8$ GeV.}
\end{figure}

The time evolution of the dark matter density is obtained by solving the Boltzmann differential equation, 
\begin{equation}\label{boltz}
 \frac{dn_{s}}{dt} = -3Hn_{s} -\braket{ \sigma_{\text{ann}}v_{\text{rel}}}
 \left [n_{s}^2-\left( n^{\text{eq}}_{s}\right )^2 \right]\,,
  \end{equation}
where $H$ stands for the Hubble expansion rate of the universe (not to be confused with the Higgs doublet denoted in section \ref{phase}), 
$ \sigma_{\text{ann}}$ is the dark matter annihilation cross section, 
$v_{\text{rel}}$  is the dark matter relative velocity and the bracket means the thermal average. 
Exploiting the {\tt micrOMEGAs} package 
\cite{Barducci:2016pcb}, we obtained the relic density. Having applied the stability conditions, 
we scanned over all DM masses in the range $10$ GeV$-10$ TeV while the only relevant coupling $\lambda_{hs}$ 
being in the interval $-2<\lambda_{hs}<2$. The result is  demonstrated in Fig. \ref{relic}. As seen in the plot, 
the dark matter particle, $s$, 
enjoys a viable space of the coupling $\lambda_{hs}$ to account for all the relic abundance observed 
in the universe by WMAP/Planck \cite{Hinshaw:2012aka,Ade:2013zuv} 
 to be $\Omega_{\text{DM}} h^2 \sim 0.12$. Furthermore, we observe from Fig. \ref{relic} that the coupling 
$\lambda_{hs}$ has is decreasing until $m_s\equiv m_{\text{DM}}=m_h/2$ and it gets larger for smaller masses afterwards. 

To calculate the DM-nucleon elastic scattering cross section, there is only one $t$-channel $ss\bar q q$ Feynman diagram which can be 
easily read from the Lagrangian.  However, one needs to use the effective operator for the interaction $ss \bar N N$ which requires the use 
of the nucleon form factors. The DM-nucleon spin-independent elastic scattering cross section is then as follows, 
\begin{equation}
 \sigma^{\text{N}}_{\text{SI}} = 
\frac{\alpha_{N}^2 \mu_{N}^2}{\pi m_{\text{DM}}^2}\,,
\end{equation}
where $\alpha_{N}$ is the effective coupling given in terms of the form factors and $\mu_{N}$ is the DM-nucleon 
reduced mass (see e.g. \cite{Ghorbani:2014gka} for more details). 
The DM-nucleon cross section can also be computed in the {\tt micrOMEGAs} package. For each set of the parameters that 
lead to the correct dark matter relic abundance, we have computed the scattering cross section. 
The result in Fig. \ref{dir} shows that the cross section saturates the LUX bound \cite{Akerib:2016vxi} 
at  $\sim m_s\equiv m_{\text{DM}}\sim 350$ GeV 
and the XENON1t limit \cite{Aprile:2017iyp} at $\sim m_{\text{DM}}\sim 600$ GeV and of course for the resonance region. 
We do not focus on the resonance region because it is excluded when we add the washout criterion into our computation as comes later on. 
We therefore conservatively can exclude the DM mass
to be $m_{\text{DM}}\gtrsim 600$ GeV if only the total relic density and the direct detection bounds are taken into consideration. 
This is consistent with the results in Ref. \cite{Athron:2017kgt} suggesting a viable DM mass  
of order $\sim 1$ TeV. Note that we have not considered all the constraints in Ref. \cite{Athron:2017kgt} 
that is why the viable DM mass seen 
in Fig. \ref{dir} has only a lower bound. 
In Fig. \ref{relic} the coupling is of order of one as 
has been mentioned in the results of the GAMBIT global fit \cite{Athron:2017kgt}.

We finally take into account the first-order electroweak phase transition conditions and the washout criterion obtained in 
Eqs. (\ref{Tc}), (\ref{globmin}), \ref{localmincon} and (\ref{vctc}). 
In Fig. \ref{dmdir}, the spin-independent DM-nucleon cross section against the DM mass is shown where all the constraints from the strongly first-order EWPT are imposed. The DM relic density condition is relaxed to take all values from $0\%$ to $100\%$ of the DM relic density observed by WMAP/Planck. As seen in Fig. \ref{dmdir}, we find regions of the parameter space that the singlet scalar covers from $0\%$ to $100\%$ of the DM relic density in a range of DM mass $\sim 6-200$ GeV. We therefore have demonstrated that before imposing any constraint from the invisible Higgs decay width, the singlet scalar covers almost all the observed relic density content of the dark matter. This is a remarkable result which has not been not pointed out in the literature. If we impose also the invisible Higgs decay bound the scalar can then cover less than $3\%$ of the relic abundance, as reported in Ref. \cite{Cline:2012hg}. Except for a small range of DM mass i.e. $6-8$ GeV, we have shown in Fig. \ref{dmdir} that all the parameter space is excluded by the XENON100 cross section bound, while in \cite{Cline:2012hg} a wide range of the DM mass evades the XENON100 upper limit. 
In fact, it is seen promptly from Fig. \ref{dmdir} that 
the smaller the fraction of the DM relic density is covered by the singlet scalar model, 
the more strongly it is excluded by the XENON100 limit. In Fig. \ref{dmdir} we have included also the updated DM-nucleon cross section bound from LUX/XENON1T.

\section{Conclusion}\label{conc}

In this article we have studied the real singlet scalar dark matter model which possesses only two additional independent parameters compared to 
the SM. The extra scalar in the model is used to simultaneously play the role of the dark matter particle and to make
the electroweak phase transition strongly first-order. By introducing a weaker first-order EWPT condition that usually is considered in the literature, we have shown that the singlet scalar model 
is capable of explaining partially or the whole dark matter relic abundance observed by the WMAP/Planck, 
and giving rise at the same time to strongly first-order EWPT. The model is excluded entirely 
by the XENON1t experiment but if we consider the XENON100 limit, the viable DM mass sits in a region of $6-8$ GeV to which the direct detection experiments are not very sensitive.

\acknowledgments
The authors would like to thank the CERN Theoretical Physics Department (CERN-TH), for the hospitality and support where
this work was finalized during their visits.

\bibliographystyle{ieeetr}
\bibliography{ref}
\end{document}